\documentclass[paper,notoc,nohyper]{JHEP}
\usepackage{epsfig}
\usepackage{amsbsy} 

\def\T{{\cal T}}
\def\C{{\cal C}}
\def\L{{\cal L}}

\def\F{{\cal F}}
\def\G{{\cal G}}

\def\Bfin{{\rm Box^6} (u, t) }
\def\Bubl{{\rm Bub}(s)}

\def\AAA{C_1}
\def\BBB{C_2}
\def\CCC{C_3}

\def\Lx{X}
\def\Ly{Y}
\def\Ls{S}
\def\Poles{{\cal P}oles}
\def\Finite{{\cal F}inite}
\def\Libx{{\rm Li}_2(x)}

\def\Licx{{\rm Li}_3(x)}
\def\Licy{{\rm Li}_3(y)}
\def\Lidx{{\rm Li}_4(x)}
\def\Lidy{{\rm Li}_4(y)}
\def\Lidz{{\rm Li}_4(z)}

\def\ttouu{\frac{t^2}{u^2}}

\def\one{}
\def\tou{\frac{t}{u}}

\def\ttoss{\frac{t^2}{s^2}}

\def\tttouuu{\frac{t^3}{u^3}}
\def\ttttouuuu{\frac{t^4}{u^4}}
\def\ttttossss{\frac{t^4}{s^4}}

\def\CA{C_A}

\def\NF{N_F}

\def\C{{\cal C}}
\def\D{{\cal D}}
\renewcommand\O[1]{{\cal O}\left(#1\right)}
\def\as{\ensuremath{\alpha_{s}}}
\def\a0{\alpha_0}

\def\Re{\mathop{\rm Re}}

\def\beq{\begin{equation}}
\def\eeq{\end{equation}}

\def\beqn{\begin{eqnarray}}
\def\eeqn{\end{eqnarray}}

\def\lq{\left[}
\def\rq{\right]}

\def\({\left(}
\def\){\right)}

\def\ket#1{|{#1}\rangle}
\def\bra#1{\langle{#1}|}
\def\braket#1#2{\langle #1 |#2 \rangle}
\def\cm{{\cal M}}

\def\MSbar{$\overline{{\rm MS}}$}

\def\bom#1{{\mbox{\boldmath $#1$}}}

\def\fs{\(-\frac{\mu^2}{s+i0}\)^\ep }
\def\ft{\(-\frac{\mu^2}{t}\)^\ep }
\def\fu{\(-\frac{\mu^2}{u}\)^\ep }

\def \ep{\epsilon}

\title{\boldmath One-loop QCD corrections to gluon-gluon scattering at 
NNLO\footnote{Work supported in part by the UK Particle Physics and Astronomy 
Research Council and by the EU Fourth Framework Programme `Training and 
Mobility of Researchers', Network `Quantum Chromodynamics and the  Deep
Structure of Elementary Particles', contract FMRX-CT98-0194 (DG 12 - MIHT).
M.E.T. acknowledges financial support from CONACyT and the CVCP. We thank the
British Council and German Academic Exchange Service for support under ARC
project 1050.} } 
\author{ E.~W.~N.~Glover$^a$ and M.~E.~Tejeda-Yeomans$^a$\\ $^a$Department of 
Physics, University of Durham,  Durham DH1 3LE,  England\\[1mm] E-mail:             \email{E.W.N.Glover@durham.ac.uk},
\email{M.E.Tejeda-Yeomans@durham.ac.uk}} 
\abstract{ 
We present the $\O{\as^4}$ virtual QCD corrections to gluon-gluon scattering 
due to the self-interference of the one-loop amplitude.  
We give analytic expressions renormalised in the \MSbar\ scheme and work in 
conventional dimensional regularisation. We write the structure of the infrared 
divergences from direct Feynman diagram evaluation in terms of the
Catani formalism for infrared divergences.
Formulae for the finite remainder are given  in terms of 
logarithms that are real in the physical region.   
These results, together with those previously obtained for
quark-quark and quark-gluon scattering complete the one-loop matrix
elements for $2 \to 2$ processes
needed for the next-to-next-to-leading order contribution to 
inclusive jet production at hadron colliders.  
}

\preprint{{DCTP/01/30}, {IPPP/01/15}, {hep-ph/0104178}}

\begin{document}

\section{Introduction}
\label{sec:intro}

To improve our understanding of jet production in current and future high
energy collider experiments we need more accurate perturbative calculations in
quantum chromodynamics (QCD). The progress of recent years means that  nowadays
a Next-to-Leading Order (NLO) perturbative calculation is standard (see for
example~\cite{EKS,jetrad}). 
Developments of efficient techniques for the evaluation  of matrix elements and
numerical algorithms that allow the virtual and bremstrahlung contributions to
be combined have made NLO perturbative calculations commonplace. However, new
improvements in detector technology and an  increase on statistics and
luminosity require more precise theoretical calculations. The reduction of
theoretical uncertainties  is crucial to refine our understanding of QCD and to
identify signatures of  possible new physics beyond the Standard Model.

The calculation of fixed next-to-next-to-leading order (NNLO) virtual
corrections for massless $2 \to 2$ processes is now an attainable task since a
vital breakthough has been made  in the calculation of two-loop master
integrals~\cite{planarA,nonplanarA,planarB,nonplanarB,
bastei3,bastei2,AGO3,xtri,AGO2}, in methods for reducing tensor integrals to master
integrals~\cite{planarB,nonplanarB,AGO3} and in the prediction  of the
structure of infrared divergences~\cite{catani}.  The first two-loop
contributions for light-like $2 \to 2$ scattering
were  addressed by  Bern, Dixon together with Kosower~\cite{bdk} and 
Ghinculov~\cite{BDG} who have calculated the
maximal-helicity-violating two-loop  amplitude for $gg \to gg$, and the QED
processes $e^+e^- \to \mu^+\mu^-$ and  $e^+e^- \to e^-e^+$, respectively. 
More recently, Gehrmann and Remiddi~\cite{gehrmann} have provided analytic 
expressions for the complete set of two-loop integrals with one off-shell leg.  
These integrals are sufficient to allow the two-loop amplitudes for three jet production in
$e^+e^-$ annihilation to be evaluated.

Recently, we provided analytical expressions for the $\O{\as^4}$ one-loop and
two-loop contribution to quark-quark scattering~\cite{qqQQ,1loopsquare,qqqq}, 
and quark-gluon  scattering~\cite{qqgg}, as well as 
crossed and time-reversed processes, in the limit where the quark mass is 
neglected. The $\O{\as^4}$ two-loop contribution
to gluon scattering is given in Ref.~\cite{gggg}. To 
complete the list of ingredients needed for 
the NNLO virtual corrections to gluon-gluon scattering, 
the interference term of one-loop amplitude 
with one-loop amplitude needs to be included. 
In this paper we present analytical formulae for this 
contribution using conventional dimensional regularisation (CDR) and 
renormalised in the \MSbar\ scheme.

The organisation of our paper is the following. We provide our notation and 
describe our method briefly in Sec. 2. In Section 3 we give analytic 
expressions for the interference of one-loop amplitude with one-loop 
amplitude.
In Sec. 3.1 we follow closely the notation used in Refs.~\cite{catani,gggg} 
to write the
infrared singularity structure in the \MSbar\ scheme for the one-loop
contribution to the NNLO calculation in terms of the one-loop bubble and
the one-loop box integral given in Appendix A. We show that the anticipated 
singularity structure agrees with our explicit calculation. Finally, in
Sec. 3.2 we give the finite contribution in terms of logarithms that have
no imaginary parts. We conclude with a brief summary of results.

\section{Notation} 
\label{sec:notation}
As in Ref.~\cite{gggg} we consider the scattering process
\begin{equation}
\label{eq:proc}
g (p_1) + g (p_2)  + g (p_3) + g (p_4) \to 0,
\end{equation}
where the gluons are all incoming with light-like momenta and carry colour indexes, $a_i$, in the adjoint representation.

Their total momentum is conserved and satisfies
$$
p_1^\mu+p_2^\mu+p_3^\mu+p_4^\mu = 0, \qquad p_i^2=0,
$$
where the associated Mandelstam variables are given by
\begin{equation}
s = (p_1+p_2)^2, \qquad t = (p_2+p_3)^2, \qquad u = (p_1+p_3)^2, 
\qquad s+t+u = 0.
\end{equation}

We treat all external gluon states in $D$ dimensions, i.e. we use CDR, and we 
renormalise the ultraviolet divergences in the \MSbar\ scheme.
In this scheme, the renormalised four point amplitude can be written as
\beqn
\label{eq:renamp}
\ket{\cm}&=& 4\pi \as \Biggl [  \ket{\cm^{(0)}}  
+ \left(\frac{\as}{2\pi}\right)  \ket{\cm^{(1)}} 
+ \left(\frac{\as}{2\pi}\right)^2\,
 \ket{ \cm^{(2)}}  + \O{\as^3} \Biggr ],\nonumber \\
\eeqn
where $\as \equiv \alpha_s(\mu^2)$ is the running coupling at renormalisation 
scale $\mu$
and the $\ket{\cm^{(i)}}$ represents the colour-space vector describing the
renormalised
$i$-loop amplitude. The dependence on both renormalisation scale $\mu$ and
renormalisation scheme is implicit.

The squared amplitude summed over spins and colours is denoted as
\beq
\D(s,t,u) = \braket{\cm}{\cm} = \sum |{\cal M}({g + g \to  g + g })|^2,
\eeq
which is symmetric under the exchange of $s$, $t$ and $u$.

Following from Eq.~(\ref{eq:renamp}), the function $\D$ can be expanded 
perturbatively to yield
\beqn
\D(s,t,u) &=& 16\pi^2\as^2 \left[
 \D^4(s,t,u)+\left(\frac{\as}{2\pi}\right) \D^6(s,t,u)
 +\left(\frac{\as}{2\pi}\right)^2 \D^8(s,t,u) +
\O{\as^{3}}\right],\nonumber \\  
\eeqn
where
\beqn
\D^4(s,t,u) &=& \braket{\cm^{(0)}}{\cm^{(0)}} 
\nonumber\\
&=& 16\,V N^2 (1-\ep)^2\(3-\frac{ut}{s^2}-\frac{us}{t^2}-\frac{st}{u^2}\),\\
\D^6(s,t,u) &=& 
\braket{\cm^{(0)}}{\cm^{(1)}}+\braket{\cm^{(1)}}{\cm^{(0)}},\\
\D^8(s,t,u) &=& \braket{\cm^{(1)}}{\cm^{(1)}} +
\braket{\cm^{(0)}}{\cm^{(2)}}+\braket{\cm^{(2)}}{\cm^{(0)}}, 
\eeqn
and $V=N^2-1$ with $N$ the number of colours.
Expressions for $\D^6$ are given in Ref.~\cite{ES} using CDR to isolate the 
infrared and ultraviolet singularities.  Also, analytic formulae for the 
two-loop contribution to $\D^8$
\begin{equation}
\D^{8\, (2 \times 0)}(s,t,u) =
\braket{\cm^{(0)}}{\cm^{(2)}}+\braket{\cm^{(2)}}{\cm^{(0)}}
\end{equation} 
are given in Ref.~\cite{gggg}.

We now present expressions for the infrared singular and finite 
contributions to $\D^8$ due to the interference of one-loop amplitudes with 
one-loop amplitudes, namely
\begin{equation}
\label{eq:self_interf}
\D^{8\, (1 \times 1)}(s,t,u) = \braket{\cm^{(1)}}{\cm^{(1)}}.
\end{equation} 
This contribution is somewhat simpler to evaluate than the two-loop 
contribution but it is  a vital part of the NNLO virtual corrections 
and we present it here for completeness. 

As in Refs.~\cite{qqQQ,1loopsquare,qqqq,qqgg,gggg}, we use 
{\tt QGRAF}~\cite{QGRAF} 
to  produce the one-loop Feynman diagrams to construct $\ket{\cm^{(1)}}$. We 
perform the internal trace (for fermion loops) over Dirac matrices in $D$ 
dimensions. The tensor integrals that emerge from this amplitude are associated 
with scalar integrals in higher dimensions and with higher powers of 
propagators~\cite{AGO3,tarasov}.

These integrals, in turn, can be reduced to master integrals in $D=4-2\ep$ with
the systematic application of integration-by-parts identities~\cite{IBP}.
At the end, all integrals arising in the one-loop amplitudes can be written
in terms of the one-loop bubble integral in $D=4-2\ep$ and the finite one-loop box 
integral in $D=6-2\ep$ (see Appendix for expansions around $\ep = 0$). 
This choice of master integrals is not unique but is useful to have a 
natural separation of the infrared poles and the finite part of the one-loop 
amplitudes.

Finally, we project by $\bra{\cm^{(1)}}$ and perform the summation over colours and
spins. 
Note that when we sum over the gluon polarisations, we ensure
that the polarisation states are transversal by the use of an axial gauge
\begin{equation}
\sum_{{\rm spins}} \ep_{i}^{\mu}\ep_{i}^{\nu\ast} = 
-  g^{\mu \nu} + \frac{n_{i}^{\mu}p_{i}^{\nu} 
+ n_{i}^{\nu}p_{i}^{\mu}}{n_{i} \cdot p_{i}} 
\end{equation}
where $p_{i}$ is the momentum, $\ep_{i}$ is the polarisation vector and 
$n_{i}$ is an arbitrary light-like 4-vector for gluon $i$. 
For simplicity, we choose 
$n_1^{\mu} = p_2^{\mu}$, $n_2^{\mu} = p_1^{\mu}$,
$n_3^{\mu} = p_4^{\mu}$ and $n_4^{\mu} = p_3^{\mu}$.

\section{One-loop contribution}
\label{sec:two}
We further decompose the one-loop contributions as a sum of two terms
\beq
\D^{8 \, (1\times 1)}(s,t,u)
 = \Poles(s,t,u)+\Finite(s,t,u).
\eeq 
$\Poles$ contains infrared singularities that will be  analytically
canceled by those occurring in radiative processes of the
same order (ultraviolet divergences are removed by renormalisation).
$\Finite$ is the remainder which is finite as $\ep \to 0$.

\subsection{Infrared Pole Structure}
\label{subsec:poles}
Following the procedure outlined in Refs.~\cite{catani,gggg}, we can write the
infrared pole structure of the one-loop contributions renormalised in the 
\MSbar\ scheme as
\begin{eqnarray}
\label{eq:poles}
\Poles &=& 2 \Re \Biggl[ -\frac{1}{2}\bra{\cm^{(0)}} {\bom I}^{(1)\dagger}(\ep){\bom I}^{(1)}(\ep) \ket{\cm^{(0)}}
\nonumber \\
&&
-\frac{\beta_0}{\epsilon} 
\,\bra{\cm^{(0)}} {\bom I}^{(1)}(\ep)  \ket{\cm^{(0)}}
+\,  \bra{\cm^{(1,un)}} {\bom I}^{(1)}(\ep)  \ket{\cm^{(0)}} \Biggr]
 \nonumber\\
\end{eqnarray}
where $\ket{\cm^{(0)}}$ is the tree amplitude and $\ket{\cm^{(1,un)}}$ is the 
unrenormalised one-loop amplitude, obtained by direct Feynman diagram 
evaluation. Also, the first coefficient of the QCD beta 
function for $N_F$ (massless) quark flavours is
\beq
\beta_0 = \frac{11 \CA - 4 T_R \NF}{6} \qquad {\rm with} \qquad   \quad\quad \CA = N, 
\qquad T_R = \frac{1}{2}.
\eeq
It is convenient to decompose $\ket{\cm^{(0)}}$ and $\ket{\cm^{(1,un)}}$ in
terms 
of $SU(N)$ matrices in the fundamental representation,
$T^{a}$,~\cite{coldeca,coldecb,coldecc,coldec1,coldec2}
so that $\ket{\cm^{(0)}}$ and $\ket{\cm^{(1,un)}}$ 
may be expressed as nine-dimensional vectors in 
colour space
\beqn
\ket{\cm^{(0)}}&=&\left(
\T_1,  \
\T_2, \
\T_3, \
\T_4,  \
\T_5,  \
\T_6,  \ 
0,\
0,\
0 \right)^T,\\
\ket{\cm^{(1,un)}}&=&\left(
\L_1,  \
\L_2,  \
\L_3,  \
\L_4,  \
\L_5,  \
\L_6,  \ 
\L_7,  \
\L_8,  \
\L_9  \right)^T,
\eeqn
where ${}^T$ indicates the transpose vector. Here the tree and loop amplitudes
$\T_i$ and $\L_i$ are the
components of $\ket{\cm^{(0)}}$ and $\ket{\cm^{(1,un)}}$
in the colour space spanned by the (non-orthogonal) basis 
\begin{eqnarray}
\label{eq:c_i}
\C_1 &=& {\rm Tr}\(T^{a_1}T^{a_2}T^{a_3}T^{a_4}\),\nonumber \\
\C_2 &=& {\rm Tr}\(T^{a_1}T^{a_2}T^{a_4}T^{a_3}\),\nonumber \\
\C_3 &=& {\rm Tr}\(T^{a_1}T^{a_4}T^{a_2}T^{a_3}\),\nonumber \\
\C_4 &=& {\rm Tr}\(T^{a_1}T^{a_3}T^{a_2}T^{a_4}\),\nonumber \\
\C_5 &=& {\rm Tr}\(T^{a_1}T^{a_3}T^{a_4}T^{a_2}\),\nonumber \\
\C_6 &=& {\rm Tr}\(T^{a_1}T^{a_4}T^{a_3}T^{a_2}\),\nonumber \\
\C_7 &=& {\rm Tr}\(T^{a_1}T^{a_2}\){\rm Tr}\(T^{a_3}T^{a_4}\),\nonumber \\
\C_8 &=& {\rm Tr}\(T^{a_1}T^{a_3}\){\rm Tr}\(T^{a_2}T^{a_4}\),\nonumber \\
\C_9 &=& {\rm Tr}\(T^{a_1}T^{a_4}\){\rm Tr}\(T^{a_2}T^{a_3}\).
\end{eqnarray}
Note that the amplitudes themselves are not required
since we compute the interference of tree and loop amplitudes directly.

In the same colour basis, the infrared-singularity operator
$\bom{I}^{(1)}(\ep)$ introduced by Catani~\cite{catani} has the form~\cite{gggg}
\begin{eqnarray}
\lefteqn{\bom{I}^{(1)}(\ep) = - \frac{e^{\ep\gamma}}{\Gamma(1-\ep)}
\left(\frac{1}{\ep^2}+\frac{\beta_0}{N\ep}\right)}
\nonumber \\
&\times&\left(
{\small
\begin{array}{ccccccccc}
N({\tt S}+{\tt T}) & 0 & 0 & 0 & 0 & 0 & ({\tt T}-{\tt U}) & 0 & ({\tt S}-{\tt U}) \\
0 & N({\tt S}+{\tt U}) & 0 & 0 & 0 & 0 & ({\tt U}-{\tt T}) & ({\tt S}-{\tt T}) & 0 \\
0 & 0 & N({\tt T}+{\tt U}) & 0 & 0 & 0 & 0 & ({\tt T}-{\tt S}) & ({\tt U}-{\tt S}) \\
0 & 0 & 0 & N({\tt T}+{\tt U}) & 0 & 0 & 0 & ({\tt T}-{\tt S}) & ({\tt U}-{\tt S}) \\
0 & 0 & 0 & 0 & N({\tt S}+{\tt U}) & 0 & ({\tt U}-{\tt T}) & ({\tt S}-{\tt T}) & 0 \\
0 & 0 & 0 & 0 & 0 & N({\tt S}+{\tt T}) & ({\tt T}-{\tt U}) & 0 & ({\tt S}-{\tt U}) \\
({\tt S}-{\tt U}) & ({\tt S}-{\tt T}) & 0 & 0 & ({\tt S}-{\tt T}) & ({\tt S}-{\tt U}) & 2N{\tt S} & 0 & 0 \\
0 & ({\tt U}-{\tt T}) & ({\tt U}-{\tt S}) & ({\tt U}-{\tt S}) & ({\tt U}-{\tt T}) & 0 & 0 & 2N{\tt U} & 0 \\
({\tt T}-{\tt U}) & 0 & ({\tt T}-{\tt S}) & ({\tt T}-{\tt S}) & 0 & ({\tt T}-{\tt U}) & 0 & 0 & 2N{\tt T}
\end{array}
}
\right)\nonumber \\
\end{eqnarray}
where
\beq
{\tt S} = \fs,\qquad\qquad {\tt T} = \ft, \qquad\qquad {\tt U} = \fu.
\eeq
The matrix $\bom{I}^{(1)}(\ep)$ acts directly as a rotation matrix on
$\ket{\cm^{(0)}}$ 
and $\ket{\cm^{(1,un)}}$ in colour space. 
Note that on expanding ${\tt S}$,
imaginary parts are generated, the sign of which is fixed by the small imaginary
part $+i0$.
Other combinations such as 
$\bra{\cm^{(0)}}\bom{I}^{(1)\dagger}(\ep)$  are obtained using the hermitian conjugate
operator $\bom{I}^{(1)\dagger}(\ep)$ 
where the only practical change is that the sign of the
imaginary part of ${\tt S}$ is reversed.

The contraction of the colour vector $\ket{X}$ with the conjugate colour vector
$\bra{Y}$ obeys the rule
\beq
\braket{Y}{X} =\sum_{\rm spins} ~\sum_{\rm colours}
 ~\sum_{i,j=1}^9~ Y_i^* \, X_j \, \C_i^*\, \C_j.
\eeq
In evaluating these contractions, we typically encounter $\sum_{\rm colours}
\C_i^*\,\C_j$ which is given by the $ij$ component of the symmetric
matrix ${\cal
C\!C}$
\begin{equation}
{\cal C\!C} = \frac{V}{16N^2}\left(
\begin{array}{ccccccccc}
\AAA & \BBB & \BBB & \BBB & \BBB & \CCC & NV & -N & NV \\
\BBB & \AAA & \BBB & \BBB & \CCC & \BBB & NV & NV & -N \\
\BBB & \BBB & \AAA & \CCC & \BBB & \BBB & -N & NV & NV \\
\BBB & \BBB & \CCC & \AAA & \BBB & \BBB & -N & NV & NV \\
\BBB & \CCC & \BBB & \BBB & \AAA & \BBB & NV & NV & -N \\
\CCC & \BBB & \BBB & \BBB & \BBB & \AAA & NV & -N & NV \\
NV & NV & -N & -N & NV & NV & N^2 V & N^2 & N^2 \\
-N & NV & NV & NV & NV & -N & N^2 & N^2 V & N^2 \\
NV & -N & NV & NV & -N & NV & N^2 & N^2 & N^2 V
\end{array}
\),
\end{equation}
with 
\begin{equation}
\AAA = N^4 - 3N^2 + 3, \qquad \BBB = 3-N^2, \qquad \CCC = 3+N^2.
\end{equation}
Similarly, we find that the interference of the tree-level amplitudes 
$\sum_{\rm spins} \T_i^* \T_j$ is 
given by ${\cal T\!T}_{ij}$, where
\begin{equation}
\label{eq:TT}
{\cal T\!T} = \frac{64(1-\ep)^2(t^2+ut+u^2)^2}{s^2t^2u^2}\ 
{\cal V}^T {\cal V},
\eeq
and the vector ${\cal V}$ is
\beq
{\cal V} = \(u, \ t, \ s,\  s,\  t,\ u ,\ 0,\ 0,\ 0\).
\eeq
Also, the interference of the tree-level amplitudes with one-loop
amplitudes
$\sum_{\rm spins} \T_i^* \L_j$ is 
given by ${\cal T\!\!L}_{ij}$, where
\begin{equation}
\label{eq:TL}
{\cal T\!\!L} = {\cal V}^T {\cal W},
\eeq
and the vector ${\cal W}$ is
\beq
{\cal W}= \Big(
\F(s,t),\  \F(s,u) ,\ \F(u,t) ,\ \F(u,t) ,\ \F(s,u) ,\ \F(s,t),\ 
\G, \ \G,\ \G \Big).
\eeq
The function $\F(s,t)$ is symmetric under the exchange of $s$ and $t$,
while $\G$ is symmetric under the exchange of any two Mandelstam invariants,
so that
\begin{eqnarray}
\F(s,t) &=& f_1(s,t,u)+f_1(t,s,u),\\
\G &=& f_2(s,t,u)+f_2(s,u,t)+f_2(t,s,u)+f_2(t,u,s)+f_2(u,s,t)+f_2(u,t,s).
\nonumber \\
\end{eqnarray}
Here $f_1$ and $f_2$ are given in terms of the one-loop box integral in 
$D=6-2\ep$ dimensions and the one-loop bubble graph in $D=4-2\ep$ (see 
Appendix~\ref{app:master_int}),
\begin{eqnarray}
f_1(s,t,u) &=& \frac{16N(1-2\ep)}{s^2t^2}\left[2(1-\ep)^2\(s^4+s^3t+st^3+t^4\) + 3(1-5\ep)s^2t^2\right]
{\rm Box}^6(s,t)\nonumber \\
&+& \frac{8\NF(1-2\ep)}{st}\left[(1-\ep)^2 \(s^2+t^2\)+\ep(1+3\ep)st\right] {\rm Box}^6(s,t) \nonumber
\\
&-& \frac{16N(1-\ep)}{s^2t^2u\ep(3-2\ep)}
\left[\(12-22\ep+12\ep^2+2\ep^3\)s^4+\(24-58\ep+50\ep^2-6\ep^3-2\ep^4\)s^3t\right. \nonumber \\
&& \qquad \left.
+\(36-99\ep+93\ep^2-24\ep^3-2\ep^4\)s^2t^2+(1-\ep)\(24-50\ep+23\ep^2\)st^3\right.\nonumber \\
&& \qquad \left.
+4(1-\ep)(1-2\ep)(3-2\ep)t^4\right]
{\rm Bub}(t)\nonumber \\
&+& \frac{16\NF}{st^2u(3-2\ep)}\left[\(4-12\ep+16\ep^2-4\ep^3\)s^3+\(3-10\ep+23\ep^2-8\ep^3\) s^2t 
\right. \nonumber \\
&&\qquad \left. 
+
\(6-15\ep+21\ep^2-8\ep^3\) st^2 + (1-\ep)\(5-6\ep+2\ep^2\)t^3 \right] {\rm
Bub}(t),\\
f_2(s,t,u) &=& \frac{32(1-2\ep)}{u^2}\left[-4(1-\ep)^2st+3(1-5\ep)u^2\right]
{\rm Box}^6(u,t)\nonumber \\
&& + \frac{32(1-\ep)}{\ep su^2}
\left[ 4(1-2\ep)(1-\ep)t^2+(8-17\ep)(1-\ep)ut\right.\nonumber \\
&& \qquad \qquad
\left.+\(6-20\ep+15\ep^2+\ep^3\)u^2
\right] 
{\rm Bub}(s). 
\end{eqnarray}

It can be easily noted that the leading infrared singularity in 
Eq.~(\ref{eq:poles}) is $\O{1/\ep^4}$.  It is a very stringent check on the
reliability of our calculation that  the pole structure obtained by computing
the Feynman diagrams directly and introducing series expansions in $\epsilon$
for the scalar master integrals  agrees with Eq.~(\ref{eq:poles}) through to
$\O{1/\ep}$.   We therefore construct the finite remainder by subtracting
Eq.~(\ref{eq:poles}) from the full result.

\subsection{Finite contributions}
\label{subsec:finite}

The finite two-loop contribution to $\D^8(s,t,u)$ is defined as 
\beq
\Finite(s,t,u) = \D^{8\, (1 \times 1)}(s,t,u) - \Poles(s,t,u),
\eeq
where we subtract the series expansions of both $\D^{8\, (1 \times 1)}(s,t,u)$ 
and $\Poles(s,t,u)$ and set $\ep \to 0$.

For convenience, we introduce the following logarithms
\begin{equation}
\label{eq:xydef1}
\Lx = \log\left(\frac{-t}{s}\right),
\qquad \Ly = \log\left(\frac{-u}{s}\right),
\qquad \Ls = \log\left(\frac{s}{\mu^2}\right),
\end{equation}
where $\mu$ is the renormalisation scale.

We choose to present our results by grouping terms 
according to the
power of the number of colours $N$ and the number 
of light quarks $\NF$, so that 
\begin{equation}
\label{eq:zi}
Finite(s,t,u) = V 
\Bigg(
N^4 A + N^2 B + N^3\NF C  + N\NF D  
+ N^2 \NF^2 E  + \NF^2 F + \frac{\NF^2}{N^2} G
\Bigg),
\end{equation}
where
\begin{eqnarray}
\label{eq:finiteA}
{A}&=&\Biggl \{ { 1\over 2}\,{}\Biggl ({\Lx^2}-{2}\,{\Lx}\,{\Ly}+{\Ly^2}+{\pi^2}\Biggr ) \Biggl ({\Lx^2}-{2}\,{\Lx}\,{\Ly}-{2}\,{\Lx}+{\Ly^2}+{2}\,{\Ly}+{\pi^2}\Biggr ) {\ttttossss}\nonumber \\ & & {}\,
+{}\Biggl ({3}\,{\Lx^4}-{4}\,{\Lx^3}\,{\Ly}-{56\over 3}\,{\Lx^3}+{6}\,{\Lx^2}\,{\Ly^2}+{20}\,{\Lx^2}\,{\Ly}-{22\over 3}\,{\Lx^2}\,{\Ls}+{10}\,{\Lx^2}\,{\pi^2}+{56\over 9}\,{\Lx^2}\nonumber \\ & & {}\,
-{4}\,{\Lx}\,{\Ly^3}-{20}\,{\Lx}\,{\Ly^2}-{4}\,{\Lx}\,{\Ly}\,{\pi^2}-{6}\,{\Lx}\,{\Ly}+{154\over 9}\,{\Lx}\,{\Ls}-{16}\,{\Lx}\,{\pi^2}+{785\over 27}\,{\Lx}+{\Ly^4}\nonumber \\ & & {}\,
+{4}\,{\Ly^3}-{22\over 3}\,{\Ly^2}\,{\Ls}+{2}\,{\Ly^2}\,{\pi^2}-{28\over 9}\,{\Ly^2}+{110\over 3}\,{\Ly}\,{\Ls}+{16}\,{\Ly}\,{\pi^2}+{721\over 9}\,{\Ly}+{242\over 9}\,{\Ls^2}\nonumber \\ & & {}\,
+{2948\over 27}\,{\Ls}+{\pi^4}+{\pi^2}+{9014\over 81}\Biggr ) {\ttoss}\nonumber \\ & & {}\,
+{}\Biggl ({4}\,{\Lx^4}+{12}\,{\Lx^3}+{4\over 3}\,{\Lx^2}\,{\Ly}-{44\over 3}\,{\Lx^2}\,{\Ls}+{16}\,{\Lx^2}\,{\pi^2}-{56\over 9}\,{\Lx^2}+{40\over 3}\,{\Lx}\,{\Ly^2}\nonumber \\ & & {}\,
+{88\over 3}\,{\Lx}\,{\Ly}\,{\Ls}+{880\over 9}\,{\Lx}\,{\Ly}+{220\over 3}\,{\Lx}\,{\Ls}+{88\over 3}\,{\Lx}\,{\pi^2}+{1442\over 9}\,{\Lx}+{4}\,{\Ly^4}-{88\over 3}\,{\Ly^2}\,{\Ls}\nonumber \\ & & {}\,
+{16}\,{\Ly^2}\,{\pi^2}-{536\over 9}\,{\Ly^2}+{116\over 3}\,{\Ly}\,{\pi^2}+{484\over 9}\,{\Ls^2}-{44\over 3}\,{\Ls}\,{\pi^2}+{5896\over 27}\,{\Ls}+{58\over 9}\,{\pi^2}+{18028\over 81}\Biggr ) {}\,{\tou}\nonumber \\ & & {}\,
+{\Lx^2}\,{}\Biggl ({\Lx^2}+{4}\,{\pi^2}\Biggr ) {}\,{\ttttouuuu}\nonumber \\ & & {}\,
+{2}\,{\Lx}\,{}\Biggl ({\Lx^3}+{2}\,{\Lx^2}+{4}\,{\Lx}\,{\pi^2}+{4}\,{\pi^2}\Biggr )  {}\,{\tttouuu}\nonumber \\ & & {}\,
+{}\Biggl ({7}\,{\Lx^4}-{8}\,{\Lx^3}\,{\Ly}-{26\over 3}\,{\Lx^3}+{12}\,{\Lx^2}\,{\Ly^2}+{88\over 3}\,{\Lx^2}\,{\Ly}-{44\over 3}\,{\Lx^2}\,{\Ls}+{24}\,{\Lx^2}\,{\pi^2}+{28\over 9}\,{\Lx^2}\nonumber \\ & & {}\,
-{8}\,{\Lx}\,{\Ly^3}-{44\over 3}\,{\Lx}\,{\Ly^2}+{88\over 3}\,{\Lx}\,{\Ly}\,{\Ls}-{8}\,{\Lx}\,{\Ly}\,{\pi^2}+{536\over 9}\,{\Lx}\,{\Ly}+{484\over 9}\,{\Lx}\,{\Ls}-{8\over 3}\,{\Lx}\,{\pi^2}\nonumber \\ & & {}\,
+{2948\over 27}\,{\Lx}+{4}\,{\Ly^4}-{88\over 3}\,{\Ly^2}\,{\Ls}+{12}\,{\Ly^2}\,{\pi^2}-{536\over 9}\,{\Ly^2}+{88\over 3}\,{\Ly}\,{\pi^2}+{484\over 9}\,{\Ls^2}-{44\over 3}\,{\Ls}\,{\pi^2}\nonumber \\ & & {}\,
+{5896\over 27}\,{\Ls}+{2}\,{\pi^4}+{10\over 9}\,{\pi^2}+{18028\over 81}\Biggr ) {}\,{\ttouu}\nonumber \\ & & {}\,
+{}\Biggl ({17\over 2}\,{\Lx^4}-{10}\,{\Lx^3}\,{\Ly}-{7\over 3}\,{\Lx^3}+{15\over 2}\,{\Lx^2}\,{\Ly^2}+{11}\,{\Lx^2}\,{\Ly}-{110\over 3}\,{\Lx^2}\,{\Ls}+{29}\,{\Lx^2}\,{\pi^2}\nonumber \\ & & {}\,
-{122\over 3}\,{\Lx^2}+{22}\,{\Lx}\,{\Ly}\,{\Ls}-{5}\,{\Lx}\,{\Ly}\,{\pi^2}+{596\over 9}\,{\Lx}\,{\Ly}+{814\over 9}\,{\Lx}\,{\Ls}+{107\over 3}\,{\Lx}\,{\pi^2}\nonumber \\ & & {}\,
+{5309\over 27}\,{\Lx}+{605\over 9}\,{\Ls^2}-{11}\,{\Ls}\,{\pi^2}+{7667\over 27}\,{\Ls}+{5\over 4}\,{\pi^4}+{113\over 18}\,{\pi^2}+{24533\over 81}\Biggr ) {}\,{\one}
\Biggr \} + \Biggl \{ u \leftrightarrow t \Biggr \}\nonumber \\
\end{eqnarray}
\begin{eqnarray}
\label{eq:finiteB}
{B}&=&\Biggl \{ { 6}\,{}\Biggl ({\Lx^2}-{2}\,{\Lx}\,{\Ly}+{\Ly^2}+{\pi^2}\Biggr ) {}\,{}\Biggl ({\Lx^2}-{2}\,{\Lx}\,{\Ly}-{2}\,{\Lx}+{\Ly^2}+{2}\,{\Ly}+{\pi^2}\Biggr ) {}\,{\ttttossss}\nonumber \\ & & 
+{}\Biggl ({72}\,{\Lx^4}-{120}\,{\Lx^3}\,{\Ly}-{356}\,{\Lx^3}+{48}\,{\Lx^2}\,{\Ly^2}+{580}\,{\Lx^2}\,{\Ly}+{156}\,{\Lx^2}\,{\pi^2}+{1280\over 3}\,{\Lx^2}\nonumber \\ & & 
+{24}\,{\Lx}\,{\Ly^3}-{404}\,{\Lx}\,{\Ly^2}-{144}\,{\Lx}\,{\Ly}\,{\pi^2}-{1184\over 3}\,{\Lx}\,{\Ly}-{392}\,{\Lx}\,{\pi^2}-{112}\,{\Lx}-{24}\,{\Ly^4}\nonumber \\ & & 
+{180}\,{\Ly^3}-{12}\,{\Ly^2}\,{\pi^2}-{32}\,{\Ly^2}+{392}\,{\Ly}\,{\pi^2}+{112}\,{\Ly}+{12}\,{\pi^4}+{12}\,{\pi^2}\Biggr ) {}\,{\ttoss}\nonumber \\ & & 
+{}\Biggl ({}-{24}\,{\Lx^4}+{144}\,{\Lx^3}\,{\Ly}+{408}\,{\Lx^3}-{48}\,{\Lx^2}\,{\Ly^2}-{272}\,{\Lx^2}\,{\Ly}+{120}\,{\Lx^2}\,{\pi^2}-{64}\,{\Lx^2}\nonumber \\ & & 
+{96}\,{\Lx}\,{\Ly^3}+{624}\,{\Lx}\,{\Ly^2}+{288}\,{\Lx}\,{\Ly}\,{\pi^2}+{2752\over 3}\,{\Lx}\,{\Ly}+{792}\,{\Lx}\,{\pi^2}+{224}\,{\Lx}\nonumber \\ & & 
+{144}\,{\Ly^2}\,{\pi^2}+{528}\,{\Ly}\,{\pi^2}+{2200\over 3}\,{\pi^2}\Biggr ) {}\,{\tou}\nonumber \\ & & 
+{12}\,{\Lx^2}\,{}\Biggl ({\Lx^2}+{4}\,{\pi^2}\Biggr ) {}\,{\ttttouuuu}\nonumber \\ & & 
+{24}\,{\Lx}\,{}\Biggl ({\Lx^3}+{2}\,{\Lx^2}+{4}\,{\Lx}\,{\pi^2}+{4}\,{\pi^2}\Biggr ) {}\,{\tttouuu}\nonumber \\ & & 
+{}\Biggl ({84}\,{\Lx^4}-{96}\,{\Lx^3}\,{\Ly}-{104}\,{\Lx^3}+{96}\,{\Lx^2}\,{\Ly^2}+{352}\,{\Lx^2}\,{\Ly}+{288}\,{\Lx^2}\,{\pi^2}+{1184\over 3}\,{\Lx^2}\nonumber \\ & & 
-{96}\,{\Lx}\,{\Ly}\,{\pi^2}-{32}\,{\Lx}\,{\pi^2}+{96}\,{\Ly^2}\,{\pi^2}+{352}\,{\Ly}\,{\pi^2}+{24}\,{\pi^4}+{1112\over 3}\,{\pi^2}\Biggr ) {}\,{\ttouu}\nonumber \\ & & 
+{}\Biggl ({42}\,{\Lx^4}+{32}\,{\Lx^3}+{66}\,{\Lx^2}\,{\Ly^2}+{416}\,{\Lx^2}\,{\Ly}+{288}\,{\Lx^2}\,{\pi^2}+{1808\over 3}\,{\Lx^2}+{84}\,{\Lx}\,{\Ly}\,{\pi^2}\nonumber \\ & & 
+{424\over 3}\,{\Lx}\,{\Ly}+{716}\,{\Lx}\,{\pi^2}+{112}\,{\Lx}+{15}\,{\pi^4}+{666}\,{\pi^2}+{48}\Biggr ) {}\,{\one}\Biggr \} + \Biggl \{ u \leftrightarrow t \Biggr \}
\end{eqnarray}
\begin{eqnarray}
\label{eq:finiteC}
{C}&=&\Biggl \{{ }-{}\Biggl ({\Lx^2}-{2}\,{\Lx}\,{\Ly}+{\Ly^2}+{\pi^2}\Biggr ) {}\,{}\Biggl ({\Lx^2}-{2}\,{\Lx}\,{\Ly}-{2}\,{\Lx}+{\Ly^2}+{2}\,{\Ly}+{\pi^2}\Biggr )  {}\,{\ttttossss}\nonumber \\ & &
+{}\Biggl ({}-{\Lx^4}+{2}\,{\Lx^3}\,{\Ly}+{73\over 6}\,{\Lx^3}-{3}\,{\Lx^2}\,{\Ly^2}-{22}\,{\Lx^2}\,{\Ly}+{19\over 6}\,{\Lx^2}\,{\Ls}-{3}\,{\Lx^2}\,{\pi^2}\nonumber \\ & &
-{185\over 18}\,{\Lx^2}+{2}\,{\Lx}\,{\Ly^3}+{22}\,{\Lx}\,{\Ly^2}+{2}\,{\Lx}\,{\Ly}\,{\pi^2}+{12}\,{\Lx}\,{\Ly}-{83\over 9}\,{\Lx}\,{\Ls}+{32\over 3}\,{\Lx}\,{\pi^2}\nonumber \\ & &
-{250\over 27}\,{\Lx}-{19\over 2}\,{\Ly^3}-{1\over 2}\,{\Ly^2}\,{\Ls}+{\Ly^2}\,{\pi^2}-{127\over 18}\,{\Ly^2}-{31\over 3}\,{\Ly}\,{\Ls}-{32\over 3}\,{\Ly}\,{\pi^2}-{242\over 9}\,{\Ly}\nonumber \\ & &
-{88\over 9}\,{\Ls^2}-{976\over 27}\,{\Ls}-{1\over 2}\,{\pi^4}-{2}\,{\pi^2}-{2752\over 81}\Biggr ) {}\,{\ttoss}\nonumber \\ & &
+{}\Biggl ({}-{4}\,{\Lx^4}-{4}\,{\Lx^3}\,{\Ly}-{27}\,{\Lx^3}+{6}\,{\Lx^2}\,{\Ly^2}+{12}\,{\Lx^2}\,{\Ly}-{\Lx^2}\,{\Ls}-{18}\,{\Lx^2}\,{\pi^2}\nonumber \\ & &
-{127\over 9}\,{\Lx^2}-{4}\,{\Lx}\,{\Ly^3}-{11}\,{\Lx}\,{\Ly^2}+{2}\,{\Lx}\,{\Ly}\,{\Ls}-{4}\,{\Lx}\,{\Ly}\,{\pi^2}-{148\over 9}\,{\Lx}\,{\Ly}-{62\over 3}\,{\Lx}\,{\Ls}\nonumber \\ & &
-{149\over 3}\,{\Lx}\,{\pi^2}-{484\over 9}\,{\Lx}+{16\over 3}\,{\Ly^2}\,{\Ls}-{2}\,{\Ly^2}\,{\pi^2}+{80\over 9}\,{\Ly^2}-{16}\,{\Ly}\,{\pi^2}-{176\over 9}\,{\Ls^2}\nonumber \\ & &
-{\Ls}\,{\pi^2}-{1952\over 27}\,{\Ls}+{\pi^4}-{247\over 9}\,{\pi^2}-{5504\over 81}\Biggr ) {}\,{\tou}\nonumber \\ & &
-{2}\,{\Lx^2}\,{}\Biggl ({\Lx^2}+{4}\,{\pi^2}\Biggr ) {}\,{\ttttouuuu}\nonumber \\ & &
-{4}\,{\Lx}\,{}\Biggl ({\Lx^3}+{2}\,{\Lx^2}+{4}\,{\Lx}\,{\pi^2}+{4}\,{\pi^2}\Biggr ) {}\,{\tttouuu}\nonumber \\ & &
+{}\Biggl ({}-{7}\,{\Lx^4}-{28\over 3}\,{\Lx^3}-{16\over 3}\,{\Lx^2}\,{\Ly}+{8\over 3}\,{\Lx^2}\,{\Ls}-{28}\,{\Lx^2}\,{\pi^2}-{52\over 3}\,{\Lx^2}+{8\over 3}\,{\Lx}\,{\Ly^2}\nonumber \\ & &
-{16\over 3}\,{\Lx}\,{\Ly}\,{\Ls}-{80\over 9}\,{\Lx}\,{\Ly}-{176\over 9}\,{\Lx}\,{\Ls}-{64\over 3}\,{\Lx}\,{\pi^2}-{976\over 27}\,{\Lx}+{16\over 3}\,{\Ly^2}\,{\Ls}+{80\over 9}\,{\Ly^2}\nonumber \\ & &
-{16\over 3}\,{\Ly}\,{\pi^2}-{176\over 9}\,{\Ls^2}+{8\over 3}\,{\Ls}\,{\pi^2}-{1952\over 27}\,{\Ls}-{40\over 3}\,{\pi^2}-{5504\over 81}\Biggr )  {}\,{\ttouu}\nonumber \\ & &
+{}\Biggl ({}-{4}\,{\Lx^4}+{2}\,{\Lx^3}\,{\Ly}-{13\over 2}\,{\Lx^3}-{3\over 2}\,{\Lx^2}\,{\Ly^2}-{2}\,{\Lx^2}\,{\Ly}+{73\over 6}\,{\Lx^2}\,{\Ls}-{15}\,{\Lx^2}\,{\pi^2}\nonumber \\ & &
-{131\over 18}\,{\Lx^2}-{4}\,{\Lx}\,{\Ly}\,{\Ls}+{\Lx}\,{\Ly}\,{\pi^2}-{97\over 9}\,{\Lx}\,{\Ly}-{269\over 9}\,{\Lx}\,{\Ls}-{94\over 3}\,{\Lx}\,{\pi^2}-{1936\over 27}\,{\Lx}\nonumber \\ & &
-{220\over 9}\,{\Ls^2}+{2}\,{\Ls}\,{\pi^2}-{2791\over 27}\,{\Ls}-{1\over 4}\,{\pi^4}-{367\over 18}\,{\pi^2}-{9337\over 81}\Biggr ) {}\,{\one}
\Biggr \} + \Biggl \{ u \leftrightarrow t \Biggr \}
\end{eqnarray}
\begin{eqnarray}
\label{eq:finiteD}
{D }&=&\Biggl \{{ }-{2}\,{}\Biggl ({\Lx^2}-{2}\,{\Lx}\,{\Ly}+{\Ly^2}+{\pi^2}\Biggr ) {}\,{}\Biggl ({\Lx^2}-{2}\,{\Lx}\,{\Ly}-{2}\,{\Lx}+{\Ly^2}+{2}\,{\Ly}+{\pi^2}\Biggr ) {}\,{\ttttossss}\nonumber \\ & & 
+{}\Biggl ({}-{4}\,{\Lx^4}+{8}\,{\Lx^3}\,{\Ly}+{75}\,{\Lx^3}-{6}\,{\Lx^2}\,{\Ly^2}-{389\over 3}\,{\Lx^2}\,{\Ly}-{8}\,{\Lx^2}\,{\pi^2}-{446\over 3}\,{\Lx^2}\nonumber \\ & & 
+{293\over 3}\,{\Lx}\,{\Ly^2}+{4}\,{\Lx}\,{\Ly}\,{\pi^2}+{424\over 3}\,{\Lx}\,{\Ly}+{232\over 3}\,{\Lx}\,{\pi^2}+{116\over 3}\,{\Lx}+{2}\,{\Ly^4}-{43}\,{\Ly^3}\nonumber \\ & & 
+{4}\,{\Ly^2}\,{\pi^2}+{22\over 3}\,{\Ly^2}-{232\over 3}\,{\Ly}\,{\pi^2}-{116\over 3}\,{\Ly}-{\pi^4}-{4}\,{\pi^2}\Biggr ) {}\,{\ttoss}\nonumber \\ & & 
+{}\Biggl ({}-{4}\,{\Lx^4}-{16}\,{\Lx^3}\,{\Ly}-{102}\,{\Lx^3}+{12}\,{\Lx^2}\,{\Ly^2}+{188\over 3}\,{\Lx^2}\,{\Ly}-{32}\,{\Lx^2}\,{\pi^2}+{44\over 3}\,{\Lx^2}\nonumber \\ & & 
-{8}\,{\Lx}\,{\Ly^3}-{380\over 3}\,{\Lx}\,{\Ly^2}-{24}\,{\Lx}\,{\Ly}\,{\pi^2}-{312}\,{\Lx}\,{\Ly}-{578\over 3}\,{\Lx}\,{\pi^2}-{232\over 3}\,{\Lx}-{4}\,{\Ly^2}\,{\pi^2}\nonumber \\ & & 
-{332\over 3}\,{\Ly}\,{\pi^2}+{2}\,{\pi^4}-{820\over 3}\,{\pi^2}\Biggr )  {}\,{\tou}\nonumber \\ & &
-{4}\,{\Lx^2}\,{}\Biggl ({\Lx^2}+{4}\,{\pi^2}\Biggr )  {}\,{\ttttouuuu}\nonumber \\ & &
-{8}\,{\Lx}\,{}\Biggl ({\Lx^3}+{2}\,{\Lx^2}+{4}\,{\Lx}\,{\pi^2}+{4}\,{\pi^2}\Biggr ) {}\,{\tttouuu}\nonumber \\ & & 
+{}\Biggl ({}-{14}\,{\Lx^4}+{8}\,{\Lx^3}-{64}\,{\Lx^2}\,{\Ly}-{56}\,{\Lx^2}\,{\pi^2}-{424\over 3}\,{\Lx^2}-{16}\,{\Lx}\,{\pi^2}-{64}\,{\Ly}\,{\pi^2}\nonumber \\ & & 
-{400\over 3}\,{\pi^2}\Biggr ) {}\,{\ttouu}\nonumber \\ & & 
+{}\Biggl ({}-{4}\,{\Lx^4}-{4}\,{\Lx^3}\,{\Ly}-{35\over 3}\,{\Lx^3}-{85}\,{\Lx^2}\,{\Ly}-{26}\,{\Lx^2}\,{\pi^2}-{206}\,{\Lx^2}-{2}\,{\Lx}\,{\Ly}\,{\pi^2}\nonumber \\ & & 
-{148\over 3}\,{\Lx}\,{\Ly}-{484\over 3}\,{\Lx}\,{\pi^2}-{116\over 3}\,{\Lx}-{1\over 2}\,{\pi^4}-{721\over 3}\,{\pi^2}-{16}\Biggr ) \,{\one}
\Biggr \} + \Biggl \{ u \leftrightarrow t \Biggr \}
\end{eqnarray}
\begin{eqnarray}
\label{eq:finiteE}
{E }&=&\Biggl \{{ 1\over 2}\,{}\Biggl ({\Lx^2}-{2}\,{\Lx}\,{\Ly}+{\Ly^2}+{\pi^2}\Biggr ) {}\,{}\Biggl ({\Lx^2}-{2}\,{\Lx}\,{\Ly}-{2}\,{\Lx}+{\Ly^2}+{2}\,{\Ly}+{\pi^2}\Biggr ){}\,{\ttttossss} \nonumber \\ & & 
+{}\Biggl ({}-{1\over 2}\,{\Lx^4}+{2}\,{\Lx^3}\,{\Ly}-{\Lx^3}-{3}\,{\Lx^2}\,{\Ly^2}+{2}\,{\Lx^2}\,{\Ly}-{1\over 3}\,{\Lx^2}\,{\Ls}-{\Lx^2}\,{\pi^2}+{32\over 9}\,{\Lx^2} \nonumber \\ & & 
+{2}\,{\Lx}\,{\Ly^3}-{2}\,{\Lx}\,{\Ly^2}+{2}\,{\Lx}\,{\Ly}\,{\pi^2}-{6}\,{\Lx}\,{\Ly}+{10\over 9}\,{\Lx}\,{\Ls}-{2\over 3}\,{\Lx}\,{\pi^2}-{22\over 27}\,{\Lx}-{1\over 2}\,{\Ly^4} \nonumber \\ & & 
+{\Ly^3}+{1\over 3}\,{\Ly^2}\,{\Ls}-{\Ly^2}\,{\pi^2}+{10\over 3}\,{\Ly^2}+{2\over 3}\,{\Ly}\,{\Ls}+{2\over 3}\,{\Ly}\,{\pi^2}+{34\over 9}\,{\Ly}+{8\over 9}\,{\Ls^2}+{80\over 27}\,{\Ls} \nonumber \\ & & 
-{1\over 2}\,{\pi^4}+{\pi^2}+{236\over 81}\Biggr ) {}\,{\ttoss}\nonumber \\ & & 
+{}\Biggl ({\Lx^4}+{6}\,{\Lx^3}-{4\over 3}\,{\Lx^2}\,{\Ly}+{2\over 3}\,{\Lx^2}\,{\Ls}+{4}\,{\Lx^2}\,{\pi^2}+{20\over 3}\,{\Lx^2}+{2\over 3}\,{\Lx}\,{\Ly^2}-{4\over 3}\,{\Lx}\,{\Ly}\,{\Ls}\nonumber \\ & & 
+{4\over 3}\,{\Lx}\,{\Ls}+{34\over 3}\,{\Lx}\,{\pi^2}+{68\over 9}\,{\Lx}+{4\over 3}\,{\Ly}\,{\pi^2}+{16\over 9}\,{\Ls^2}+{2\over 3}\,{\Ls}\,{\pi^2}+{160\over 27}\,{\Ls}\nonumber \\ & & 
+{22\over 3}\,{\pi^2}+{472\over 81}\Biggr ) {}\,{\tou}\nonumber \\ & & 
+{\Lx^2}\,{}\Biggl ({\Lx^2}+{4}\,{\pi^2}\Biggr )  {}\,{\ttttouuuu}\nonumber \\ & &
+{2}\,{\Lx}\,{}\Biggl ({\Lx^3}+{2}\,{\Lx^2}+{4}\,{\Lx}\,{\pi^2}+{4}\,{\pi^2}\Biggr ) \nonumber \\ & & {}\,{\tttouuu}+{}\Biggl ({2}\,{\Lx^4}+{6}\,{\Lx^3}+{8}\,{\Lx^2}\,{\pi^2}+{62\over 9}\,{\Lx^2}+{16\over 9}\,{\Lx}\,{\Ls}+{12}\,{\Lx}\,{\pi^2}+{80\over 27}\,{\Lx}+{16\over 9}\,{\Ls^2}\nonumber \\ & &
+{160\over 27}\,{\Ls}+{44\over 9}\,{\pi^2}+{472\over 81}\Biggr ) {}\,{\ttouu}\nonumber \\ & & 
+{}\Biggl ({1\over 2}\,{\Lx^4}-{\Lx^3}\,{\Ly}+{4\over 3}\,{\Lx^3}+{3\over 4}\,{\Lx^2}\,{\Ly^2}-{\Lx^2}\,{\Ls}+{3\over 2}\,{\Lx^2}\,{\pi^2}+{28\over 9}\,{\Lx^2}-{1\over 2}\,{\Lx}\,{\Ly}\,{\pi^2}\nonumber \\ & &
+{5\over 9}\,{\Lx}\,{\Ly}+{22\over 9}\,{\Lx}\,{\Ls}+{14\over 3}\,{\Lx}\,{\pi^2}+{218\over 27}\,{\Lx}+{20\over 9}\,{\Ls^2}+{254\over 27}\,{\Ls}+{1\over 8}\,{\pi^4}\nonumber \\ & &
+{37\over 9}\,{\pi^2}+{1049\over 81}\Biggr )  {}\,{\one}
\Biggr \} + \Biggl \{ u \leftrightarrow t \Biggr \}
\end{eqnarray}
\begin{eqnarray}
\label{eq:finiteF}
{F }&=&\Biggl \{{ }-{}\Biggl ({\Lx^2}-{2}\,{\Lx}\,{\Ly}+{\Ly^2}+{\pi^2}\Biggr ) {}\,{}\Biggl ({\Lx^2}-{2}\,{\Lx}\,{\Ly}-{2}\,{\Lx}+{\Ly^2}+{2}\,{\Ly}+{\pi^2}\Biggr )  {}\,{\ttttossss}\nonumber \\ & &
 +{}\Biggl ({\Lx^4}-{4}\,{\Lx^3}\,{\Ly}-{2}\,{\Lx^3}+{6}\,{\Lx^2}\,{\Ly^2}+{14\over 3}\,{\Lx^2}\,{\Ly}+{2}\,{\Lx^2}\,{\pi^2}+{6}\,{\Lx^2}-{4}\,{\Lx}\,{\Ly^3}\nonumber \\ & & 
 -{14\over 3}\,{\Lx}\,{\Ly^2}-{4}\,{\Lx}\,{\Ly}\,{\pi^2}+{4\over 3}\,{\Lx}\,{\Ly}-{4\over 3}\,{\Lx}\,{\pi^2}+{4\over 3}\,{\Lx}+{\Ly^4}+{2}\,{\Ly^3}+{2}\,{\Ly^2}\,{\pi^2}-{22\over 3}\,{\Ly^2}\nonumber \\ & & 
 +{4\over 3}\,{\Ly}\,{\pi^2}-{4\over 3}\,{\Ly}+{\pi^4}-{2}\,{\pi^2}\Biggr )  {}\,{\ttoss}\nonumber \\ & &
 +{}\Biggl ({}-{2}\,{\Lx^4}-{4}\,{\Lx^3}-{8\over 3}\,{\Lx^2}\,{\Ly}-{8}\,{\Lx^2}\,{\pi^2}-{44\over 3}\,{\Lx^2}+{8\over 3}\,{\Lx}\,{\Ly^2}+{80\over 3}\,{\Lx}\,{\Ly}-{28\over 3}\,{\Lx}\,{\pi^2}\nonumber \\ & &
 -{8\over 3}\,{\Lx}+{8\over 3}\,{\Ly}\,{\pi^2}+{16}\,{\pi^2}\Biggr )  {}\,{\tou}\nonumber \\ & &
 -{2}\,{\Lx^2}\,{}\Biggl ({\Lx^2}+{4}\,{\pi^2}\Biggr )  {}\,{\ttttouuuu}\nonumber \\ & &
 -{4}\,{\Lx}\,{}\Biggl ({\Lx^3}+{2}\,{\Lx^2}+{4}\,{\Lx}\,{\pi^2}+{4}\,{\pi^2}\Biggr )  {}\,{\tttouuu}\nonumber \\ & &
 +{}\Biggl ({}-{4}\,{\Lx^4}-{12}\,{\Lx^3}-{16}\,{\Lx^2}\,{\pi^2}-{4\over 3}\,{\Lx^2}-{24}\,{\Lx}\,{\pi^2}+{8\over 3}\,{\pi^2}\Biggr )  {}\,{\ttouu}\nonumber \\ & &
 +{}\Biggl ({}-{\Lx^4}+{2}\,{\Lx^3}\,{\Ly}-{4\over 3}\,{\Lx^3}-{3\over 2}\,{\Lx^2}\,{\Ly^2}+{2}\,{\Lx^2}\,{\Ly}-{3}\,{\Lx^2}\,{\pi^2}+{46\over 3}\,{\Lx^2}\nonumber \\ & &
 +{\Lx}\,{\Ly}\,{\pi^2}+{2}\,{\Lx}\,{\Ly}+{4\over 3}\,{\Lx}\,{\pi^2}-{4\over 3}\,{\Lx}-{1\over 4}\,{\pi^4}+{58\over 3}\,{\pi^2}-{8}\Biggr ) {}\,{\one}
\Biggr \} + \Biggl \{ u \leftrightarrow t \Biggr \}
\end{eqnarray}
\begin{eqnarray}
\label{eq:finiteG}
{G}&=&\Biggl \{{ 3}\,{}\Biggl ({\Lx^2}-{2}\,{\Lx}\,{\Ly}+{\Ly^2}+{\pi^2}\Biggr )  {}\,{}\Biggl ({\Lx^2}-{2}\,{\Lx}\,{\Ly}-{2}\,{\Lx}+{\Ly^2}+{2}\,{\Ly}+{\pi^2}\Biggr )  {}\,{\ttttossss}\nonumber \\ & &
-{3}\,{}\Biggl ({\Lx^2}-{2}\,{\Lx}\,{\Ly}+{2}\,{\Lx}+{\Ly^2}-{2}\,{\Ly}+{\pi^2}-{2}\Biggr ) \,{}\Biggl ({\Lx^2}-{2}\,{\Lx}\,{\Ly}-{2}\,{\Lx}+{\Ly^2}+{2}\,{\Ly}+{\pi^2}\Biggr ) \,{\ttoss}\nonumber \\ & & {}
+{}\Biggl ({6}\,{\Lx^4}+{24}\,{\Lx^3}+{24}\,{\Lx^2}\,{\pi^2}+{36}\,{\Lx^2}+{48}\,{\Lx}\,{\pi^2}+{24}\,{\Lx}+{24}\,{\pi^2}\Biggr )  {}\,{\tou}\nonumber \\ & &
+{6}\,{\Lx^2}\,{}\Biggl ({\Lx^2}+{4}\,{\pi^2}\Biggr )  {}\,{\ttttouuuu}\nonumber \\ & &
+{12}\,{\Lx}\,{}\Biggl ({\Lx^3}+{2}\,{\Lx^2}+{4}\,{\Lx}\,{\pi^2}+{4}\,{\pi^2}\Biggr ) {}\,{\tttouuu}\nonumber \\ & & 
+{}\Biggl ({12}\,{\Lx^4}+{36}\,{\Lx^3}+{48}\,{\Lx^2}\,{\pi^2}+{36}\,{\Lx^2}+{72}\,{\Lx}\,{\pi^2}+{24}\,{\pi^2}\Biggr )  {}\,{\ttouu}\nonumber \\ & &
+{}\Biggl ({3}\,{\Lx^4}-{6}\,{\Lx^3}\,{\Ly}+{6}\,{\Lx^3}+{9\over 2}\,{\Lx^2}\,{\Ly^2}+{9}\,{\Lx^2}\,{\pi^2}+{6}\,{\Lx^2}-{3}\,{\Lx}\,{\Ly}\,{\pi^2}+{6}\,{\Lx}\,{\Ly}\nonumber \\ & &
+{12}\,{\Lx}\,{\pi^2}+{12}\,{\Lx}+{3\over 4}\,{\pi^4}+{6}\,{\pi^2}+{24}\Biggr ) {}\,{\one}
\Biggr \} + \Biggl \{ u \leftrightarrow t \Biggr \}
\end{eqnarray}
We see that although we expect the finite piece to contain polylogarithms,
they are all predicted by the infrared 
singular structure and are obtained by
expanding Eq.~(\ref{eq:poles}) through to $\O{\ep}$.  This is because
the polylogarithms 
appear as the $\O{\ep}$ and $\O{\ep^2}$ terms in the expansion of the box
integral in $D=6$ and must be multiplied by an infrared singular term to
contribute at $\O{1}$.  At $\O{1}$, the interference of one
box graph with another only collects the $\O{1}$ terms in each and therefore
yields only logarithms.

\section{Summary}
\label{sec:conc}

In this paper we presented analytic expressions for the $\O{\as^4}$ QCD
corrections to the $2 \to 2$ gluon-gluon scattering process due to the
self-interference of the one-loop amplitude in the \MSbar\ scheme.  Throughout
we employed conventional dimensional regularisation. 

The renormalised matrix elements are infrared divergent and contain poles down
to $\O{1/\ep^4}$. The pole structure of the one-loop contribution is described
by Eq.~(\ref{eq:poles}) using the formalism of Catani~\cite{catani} while
analytic formulae for  the finite part according to the colour decomposition of
Eq.~(\ref{eq:zi}) are given in Eqs.~(\ref{eq:finiteA}) to~(\ref{eq:finiteG}). 
The interference of the tree-level diagrams with the one-loop graphs are 
expressed in terms of the one-loop bubble graph in $D=4-2\ep$ dimensions  and
the one-loop box graph in $D=6-2\ep$ dimensions for which series expansions
around $\epsilon = 0$ are provided in Appendix~\ref{app:master_int}.

The results presented here, together with those previously computed for
quark-quark scattering~\cite{qqQQ,qqqq,1loopsquare}, quark-gluon
scattering~\cite{qqgg} and gluon-gluon scattering~\cite{gggg} form a complete
set of virtual matrix elements for parton-parton scattering at $\O{\as^4}$.
They are vital ingredients for the next-to-next-to-leading order predictions
for jet cross sections in hadron-hadron collisions.   

The next step is to combine these matrix elements with the tree-level $2 \to
4$~\cite{6g,4g2q,2g4q,6q}, the one-loop $2 \to 3$~\cite{5g,3g2q,1g4q}
contributions making sure that the infrared singularities are cancelled.  This
is a challenging task and a systematic procedure needs to be established.
However, recent progress in determining the singular limits of tree-level
matrix elements when two particles are unresolved~\cite{tc,ds} and the soft and
collinear limits of one-loop amplitudes~\cite{sone,coldec2,cone}, together with
the analytic cancellation of the infrared singularities in the somewhat simpler
case of $e^+e^- \to {\rm photon} + {\rm jet}$ at next-to-leading
order~\cite{aude}, suggest that the technical problems may soon be solved for
generic $2 \to 2$ scattering processes.

Initial state radiation complicates the issue and the colinear singularities 
from the incoming partons must be factorised into the parton density functions.
The evolution of the parton density functions must also be known to an accuracy
matching  the hard scattering matrix element and requires knowledge of the
three-loop splitting functions.  Using the existing three-loop order 
moments~\cite{moms1,moms2,Gra1}, Van Neerven and Vogt have provided accurate
parameterisations of the splitting functions in $x$-space~\cite{NV,NVplb} which
are now starting to be implemented in the global analyses~\cite{MRS}.

Finally, and most importantly for phenomenological applications, an  extension
of the numerical next-to-leading order two jet programs~\cite{EKS,jetrad} to
next-to-next-to-leading order must be developed. We note that 
Refs.~\cite{trocsanyi,kilgore} have provided next-to-leading order programs for
three jet production that form a natural starting point. We are therefore
optimistic that these issues will shortly be resolved thereby enabling 
next-to-next-to-leading order QCD estimates of jet production in hadron
collisions.

\section*{Acknowledgements}

M.E.T. acknowledges financial support from CONACyT and the CVCP.  We
gratefully acknowledge the support of the British Council and German Academic
Exchange Service under ARC project 1050.
This work was supported in part by the EU Fourth Framework Programme
`Training and Mobility of Researchers', Network `Quantum Chromodynamics and
the Deep Structure of Elementary Particles', contract FMRX-CT98-0194
(DG-12-MIHT).

\appendix
\section{One-loop master integrals}
\label{app:master_int}
In this appendix, we list the expansions for the one-loop box integrals in
$D=6-2\ep$.
We remain in the physical region $s>0$, $u,t < 0$, 
and write coefficients in terms of logarithms and polylogarithms that are
real in this domain.  More precisely, we use the notation of 
Eq.~(\ref{eq:xydef1}) to define the arguments of the logarithms and the 
polylogarithms are defined as 
\begin{eqnarray}
 {\rm Li}_n(w) &=& \int_0^w \frac{dt}{t} {\rm Li}_{n-1}(t) \qquad {\rm ~for~}
 n=2,3,4\nonumber \\
 {\rm Li}_2(w) &=& -\int_0^w \frac{dt}{t} \log(1-t).
\label{eq:lidef}
\end{eqnarray} 
Using the standard polylogarithm identities~\cite{kolbig},
we retain the polylogarithms with arguments $x$, $1-x$ and
$(x-1)/x$, where
\begin{equation}
\label{eq:xydef}
x = -\frac{t}{s}, \qquad y = -\frac{u}{s} = 1-x, \qquad z=-\frac{u}{t} = \frac{x-1}{x}.
\end{equation}

We find that the box integrals have the expansion
\begin{eqnarray}
\Bfin &=& \frac{ e^{\ep\gamma}
\Gamma  \left(  1+\epsilon \right)  \Gamma  
\left( 1-\epsilon \right) ^2 
 }{ 2s\Gamma  \left( 1-2 \epsilon  \right)   \left( 1-2 \epsilon  \right) } 
 \left(\frac{\mu^2}{s} \right)^{\ep}
  \Biggl\{
 \frac{1}{2}\lq\(\Lx-\Ly\)^2+\pi^2 \rq\nonumber \\
&& 
 +2\ep \lq
 \Licx-\Lx\Libx-\frac{1}{3}\Lx^3-\frac{\pi^2}{2}\Lx \rq
\nonumber \\
&& 
-2\ep^2\Bigg[
\Lidx+\Ly\Licx-\frac{1}{2}\Lx^2\Libx-\frac{1}{8}\Lx^4-\frac{1}{6}\Lx^3\Ly+\frac{1}{4}\Lx^2\Ly^2\nonumber
\\
&&\qquad
\qquad-\frac{\pi^2}{4}\Lx^2-\frac{\pi^2}{3}\Lx\Ly-\frac{\pi^4}{45}\Bigg]
+ ( u \leftrightarrow t) \Biggr\} + \O{\ep^3},
\end{eqnarray}
and
\begin{eqnarray}
\label{eq:boxst}
{\rm Box}^6(s, t)&=&\frac{e^{\ep\gamma}\Gamma(1+\ep) \Gamma(1-\ep)^2}
{2 u\Gamma(1-2\ep)(1-2\ep)}\,\fu  \Biggl\{ \left(\Lx^2 +2 i\pi
\Lx\right)\nonumber \\
&&+\ep \Biggl[
\left(-2\Licx+2 \Lx \Libx  -\frac{2}{3} \Lx^3+2 \Ly \Lx^2-\pi^2 \Lx+2 \zeta_3
\right)\nonumber \\
&& \qquad \qquad +i\pi\left(2 \Libx +4 \Ly \Lx-\Lx^2-\frac{\pi^2}{3}\right)
\Biggr]\nonumber \\
&&+\ep^2 \Bigg[
\Biggl(2\Lidz+2\Lidy-2\Ly \Licx-2\Lx \Licy+(2\Lx\Ly-X^2-\pi^2)\Libx
\nonumber \\
&&\qquad+\frac{1}{3}\Lx^4-\frac{5}{3}\Lx^3\Ly+\frac{3}{2}\Lx^2\Ly^2+\frac{2}{3}\pi^2\Lx^2-2\pi^2\Lx\Ly+2\Ly\zeta_3+\frac{1}{6}\pi^4\Biggr)
\nonumber \\
&&\qquad \qquad + i\pi \biggl(
-2\Licx-2\Licy+2\Ly\Libx+\frac{1}{3}\Lx^3-2\Lx^2\Ly+3\Lx\Ly^2\nonumber \\
&& \qquad \qquad \qquad \qquad -\frac{\pi^2}{3}\Ly+2\zeta_3
\biggr)\Bigg] \Biggr\} + \O{\ep^3}.
\end{eqnarray}
${\rm Box}^6(s,u)$ is obtained from Eq.~(\ref{eq:boxst}) by exchanging $u$ and
$t$.

Finally, the one-loop bubble integral in $D=4-2 \epsilon$ dimensions 
is given by
\begin{equation} 
 \Bubl =\frac{ e^{\ep\gamma}\Gamma  \left(  1+\epsilon \right)  \Gamma  \left( 1-\epsilon \right) ^2 
 }{ \Gamma  \left( 2-2 \epsilon  \right)  \epsilon   } \fs.
\end{equation}


\begin{thebibliography}{99}
\bibitem{EKS} S.D. Ellis, Z. Kunzst and D.E. Soper, 
Phys. Rev. Lett. {\bf 69} (1992) 3615 [hep-ph/9208249]. 
\bibitem{jetrad} W.T. Giele, E.W.N. Glover and D.A. Kosower,
Phys. Rev. {\bf D52} (1995) 1486 [hep-ph/9412338].
\bibitem{planarA}V.A. Smirnov, Phys. Lett. {\bf B460} (1999) 397
[hep-ph/9905323].
\bibitem{nonplanarA}J.B. Tausk, Phys. Lett. {\bf B469} (1999) 225
[hep-ph/9909506].
\bibitem{planarB}V.A. Smirnov and O.L. Veretin, Nucl. Phys. {\bf B566} (2000)
469 [hep-ph/9907385].
\bibitem{nonplanarB}C. Anastasiou, T. Gehrmann, C. Oleari,
E. Remiddi and J.B. Tausk,  Nucl. Phys. {\bf B580} (2000) 577  [hep-ph/0003261].
\bibitem{bastei3}T. Gehrmann and E. Remiddi, Nucl. Phys. Proc. Suppl. {\bf 89}
(2000) 251 [hep-ph/0005232].
\bibitem{bastei2}C. Anastasiou, J.B. Tausk and M.E. Tejeda-Yeomans, 
Nucl. Phys. Proc. Suppl. {\bf 89} (2000) 262   [hep-ph/0005328].
\bibitem{AGO3}C. Anastasiou, E.W.N. Glover and   C. Oleari,
 Nucl. Phys. {\bf B575} (2000) 416,  Erratum-ibid.\ {\bf B585} (2000) 763 
[hep-ph/9912251].
\bibitem{xtri}R.J. Gonsalves, Phys. Rev. {\bf D28} (1983) 1542;\\
G. Kramer and B. Lampe, J. Math. Phys. {\bf 28} (1987) 945.
\bibitem{AGO2}C. Anastasiou, E.W.N. Glover and   C. Oleari,
 Nucl. Phys. {\bf B572} (2000) 307 [hep-ph/9907494].
\bibitem{catani}S. Catani, Phys. Lett. {\bf B427} (1998) 161 [hep-ph/9802439].
\bibitem{bdk}Z. Bern, L. Dixon and D.A. Kosower, JHEP {\bf 0001} (2000) 027
[hep-ph/0001001].
\bibitem{BDG}Z. Bern, L. Dixon and A. Ghinculov,  Phys. Rev. {\bf D63}
(2001) 053007  [hep-ph/0010075].
\bibitem{gehrmann}T.K. Gehrmann and E. Remiddi, 
Nucl. Phys. {\bf B580} (2000) 485 [hep-ph/9912329]; hep-ph/0008287; hep-ph/0101124.
 
\bibitem{qqQQ} C. Anastasiou, E.W.N. Glover, C. Oleari and M.E.
Tejeda-Yeomans, hep-ph/0010212.
\bibitem{1loopsquare} C. Anastasiou, E.W.N. Glover, C. Oleari and M.E.
Tejeda-Yeomans, hep-ph/0012007.
\bibitem{qqqq} C. Anastasiou, E.W.N. Glover, C. Oleari and M.E.
Tejeda-Yeomans, hep-ph/0011094.
\bibitem{qqgg} C. Anastasiou, E.W.N. Glover, C. Oleari and M.E.
Tejeda-Yeomans, hep-ph/0101304.
\bibitem{gggg} E.W.N. Glover, C. Oleari and M.E.Tejeda-Yeomans, 
hep-ph/0102201.
\bibitem{ES} R.K. Ellis and J.C. Sexton, Nucl. Phys. {\bf B269} (1986) 445.

\bibitem{QGRAF}P. Nogueira, J. Comput. Phys. {\bf 105} (1993) 279. 

\bibitem{tarasov} O.V. Tarasov, Phys.Rev.{\bf D54} (1996) 6479 
[hep-ph/9606018], Nucl.Phys.{\bf B502} (1997) 455 [hep-ph/9703319].
\bibitem{IBP}K.G.~Chetyrkin, A.L.~Kataev and F.V.~Tkachov, 
Nucl. Phys. {\bf B174} (1980) 345 ;\\
K.G.~Chetyrkin and F.V.~Tkachov, Nucl. Phys. {\bf B192} (1981) 159.


\bibitem{coldeca} P. Cvitanovic, P.G. Lauwers and P.N. Scharbach, Nucl. Phys.
{\bf B186} (1981) 165.
\bibitem{coldecb}
F.A. Berends and W.T. Giele, Nucl. Phys. {\bf B294} (1987) 700;\\
M. Mangano, S.J. Parke and Z. Xu, Nucl. Phys. {\bf B298} (1988) 653.
\bibitem{coldecc}
D.A. Kosower, B.-H. Lee and V.P. Nair, Phys. Letts. {\bf B201} (1988) 85;\\
M. Mangano, Nucl. Phys. {\bf B309} (1988) 461;\\
D. Zeppenfeld, Int. J. Mod. Phys. {\bf A3} (1988) 2175;\\
D.A. Kosower, Nucl. Phys. {\bf B315} (1989) 391.
\bibitem{coldec1} Z. Bern and D.A. Kosower, Nucl. Phys. {\bf B362} (1991)
389;\\ 
Z. Bern, L. Dixon and D.A. Kosower, Ann. Rev. Nucl. Part. Sci.
{\bf 46} (1996) 109 [hep-ph/9602280].
\bibitem{coldec2}
Z. Bern, L. Dixon, D.C. Dunbar and D.A. Kosower, Nucl. Phys. {\bf B425} (1994)
217 [hep-ph/9403226].



\bibitem{6g} J.F. Gunion and J. Kalinowski, Phys. Rev. {\bf D34} (1986) 2119;\\
S.J. Parke and T.R. Taylor, Nucl. Phys. {\bf B269} (1986) 410;\\
F.A. Berends and W.T. Giele, Nucl. Phys. {\bf B294} (1987) 700;\\
M. Mangano, S.J. Parke and Z. Xu, Nucl. Phys. {\bf B298} (1988) 653.
\bibitem{4g2q} Z. Kunszt, Nucl. Phys. {\bf B271} (1986) 333;\\
S.J. Parke and T.J. Taylor, Phys. Rev. {\bf D35} (1987) 313.
\bibitem{2g4q}J.F. Gunion and Z. Kunszt, Phys. Lett. {\bf 159B} (1985) 167.
\bibitem{6q}J.F. Gunion and Z. Kunszt, Phys. Lett. {\bf 176B} (1986) 163.
\bibitem{5g}
Z. Bern, L. Dixon, D.A. Kosower, Phys. Rev. Lett
{\bf 70} (1993) 2677  [hep-ph/9302280].
\bibitem{3g2q}
Z. Bern, L. Dixon and D.A. Kosower, Nucl. Phys.
{\bf B437} (1995) 259  [hep-ph/9409393].
\bibitem{1g4q}
Z. Kunszt, A. Signer and Z. Tr\'ocs\'anyi, 
Phys.\ Lett.\ {\bf B336} (1994) 529  [hep-ph/9405386].


\bibitem{tc}J.M. Campbell and E.W.N. Glover, Nucl. Phys. {\bf B527} (1998)
264 [hep-ph/9710255];\\
S. Catani and M. Grazzini, Phys. Lett. {\bf 446B} (1999) 143 
[hep-ph/9810389];\\
S. Catani and M. Grazzini, Nucl. Phys. {\bf B570} (2000) 287 [hep-ph/9908523];\\
V. Del Duca, A. Frizzo and F. Maltoni, Nucl. Phys. {\bf B568} (2000) 211
[hep-ph/9909464].
\bibitem{ds}F.A. Berends and W.T. Giele, Nucl. Phys. {\bf B313} (1989) 595;\\
S. Catani, in Proceedings of the workshop on {\em New Techniques for 
Calculating Higher Order QCD Corrections}, report ETH-TH/93-01, Zurich (1992).
\bibitem{sone}Z. Bern, V. Del Duca and C.R. Schmidt, Phys. Lett. {\bf 445B}
(1998) 168 [hep-ph/9810409];\\
Z. Bern, V. Del Duca, W.B. Kilgore and C.R. Schmidt, Phys. Rev. {\bf D60} (1999)
116001 [hep-ph/9903516];\\
S. Catani and M. Grazzini, Nucl. Phys. {\bf B591} (2000) 435 [hep-ph/0007142].
\bibitem{cone}
D.A. Kosower, Nucl. Phys. {\bf B552} (1999) 319;\\
D.A. Kosower and P. Uwer, Nucl. Phys. {\bf B563} (1999) 477 [hep-ph/9903515].


\bibitem{aude} A. Gehrmann-De Ridder and E.W.N. Glover, Nucl. Phys. {\bf B517}
(1998) 269 [hep-ph/9707224].
\bibitem{moms1}  S.A. Larin, T. van Ritbergen and J.A.M. Vermaseren,
                 Nucl.\ Phys.\ {\bf B427} (1994) 41;\\
                 S.A. Larin, P. Nogueira, T. van Ritbergen and J.A.M.
                 Vermaseren, Nucl.\ Phys.\ {\bf B492} (1997) 338
		 [hep-ph/9605317].
\bibitem{moms2}  A. Retey and J.A.M. Vermaseren, hep-ph/0007294.

\bibitem{Gra1}   J.A. Gracey, Phys.\ Lett.\ {\bf B322} (1994) 141
[hep-ph/9401214].
\bibitem{NV}    W.L. van Neerven and A. Vogt, Nucl.\ Phys.\ {\bf B568}
                 (2000) 263 [hep-ph/9907472]; Nucl. Phys. {\bf 588} (2000) 345
		 [hep-ph/0006154].
		 
\bibitem{NVplb}  W.L. van Neerven and A. Vogt, Phys.\ Lett.\ {\bf B490}
                 (2000) 111 [hep-ph/0007362].

\bibitem{MRS} A.D. Martin, R.G. Roberts, W.J. Stirling and R.S. Thorne,
Eur.\ Phys.\ J.\ {\bf C18} (2000) 117 [hep-ph/0007099].
\bibitem{trocsanyi} Z. Trocsanyi, Phys.~Rev.~Lett. {\bf 77} (1996) 2182
[hep-ph/9610499].
\bibitem{kilgore} W.B.~Kilgore and W.T.~Giele, Phys.~Rev. {\bf D55} (1997)
7183 [hep-ph/9610433]; hep-ph/0009193. 
\bibitem{kolbig}
K.S.~K\"olbig, J.A.~Mignaco and E.~Remiddi,
B.I.T.\ {\bf 10} (1970) 38.


\end{thebibliography}
\end{document}